\documentclass[onecolumn,aps,prd,preprintnumbers,showpacs,superscriptaddress,nofootinbib,amsmath,amssymb,floats,floatfix,showkeys,notitlepage,longbibliography]{revtex4-1}
\usepackage{graphicx}	
\usepackage{amssymb}
\usepackage{dcolumn}
\usepackage{mathtools}
\usepackage{amsmath}
\usepackage{xcolor}
\usepackage{color}
\usepackage{theorem}
\usepackage{subfigure, rotating, bm, array}
\usepackage[pagebackref=false, colorlinks=true]{hyperref}
\hypersetup{linkcolor=blue, citecolor=blue,urlcolor=blue} 

\usepackage[english]{babel}           

\begin{document}

\title{Non-singular black hole by gravitational decoupling and some thermodynamic properties}

\author{Maxim Misyura}
\affiliation{Department of High Energy and Elementary Particles Physics, Saint Petersburg State University, University Emb. 7/9, Saint Petersburg, 199034, Russia}
\email{max.misyura94@gmail.com}
\author{Angel Rincon}
\affiliation{Departamento de Física Aplicada, Universidad de Alicante, Campus de San Vicente del Raspeig, E-03690 Alicante, Spain}
\email{angel.rincon@ua.es} 
\author{Vitalii Vertogradov}
\affiliation{Physics Department, Herzen State Pedagogical University of Russia, 48 Moika Emb., Saint Petersburg 191186, Russia
SPb branch of SAO RAS, 65 Pulkovskoe Rd, Saint Petersburg 196140, Russia}
\email{vdvertogradov@gmail.com}

\date{\today}

\begin{abstract}
Gravitational decoupling allows to obtain new solutions of general relativity. In this paper, we obtain new solutions of the Einstein field equations which describe non-singular black holes. We consider Hayward and Bardeen regular black holes as seed spacetimes and apply gravitational decoupling to obtain a new non-singular solution. We show that anisotropic energy-momentum tensor can spoil the regularity condition in the centre of a black hole. We solve the Einstein field equation and obtain new solutions that possess a de Sitter core and have Schwarzschild behaviour in infinity. We also analyse the thermodynamic properties of the obtained solutions.

\end{abstract}
\maketitle

\section{Introduction}

Black holes are considered the simplest solution of the Einstein field equations and they are usually compared with the hydrogen atom in quantum mechanics given its simplicity and because such objects are quite useful to learn about the physics in the corresponding scale. 
Although the concept of black holes has a rich historical background, Schwarzschild's seminal example, presented more than a century ago \cite{Schwarzschild:1916uq}, stands as the first representative example. The identification of black holes served to highlight the inadequacies of Newtonian physics in explaining gravity and underscored the profound consequences of Einstein's general theory of relativity. Even in scenarios without matter, Einstein's equations yield non-trivial solutions, exemplified by black holes, whose properties deviate substantially from those of a flat Minkowski spacetime.
The fascination of black holes comes from the intricate interplay between classical and quantum physics, which is essential to their understanding. A key milestone in black hole physics came with Stephen Hawking's seminal contributions \cite{Hawking:1974rv,Hawking:1975vcx}, which clarified the phenomenon of radiation emission from black hole event horizons. This pioneering discovery has effectively transformed black holes into a critical experimental arena, providing a unique laboratory setting for probing and gaining insight into the inherent complexities of gravitational theories.
The interior of black holes is still a conceptual problem due to the presence of singularities, which are expected to occur under certain conditions \cite{Hawking:1973uf}. 
Indeed, the classical solution of Einstein's field equations presents both future singularities \cite{bib:penrose} and past singularities \cite{Hawking:1965mf,Hawking:1966sx,Hawking:1966jv,Hawking:1967ju}. These singularities are usually hidden behind an event horizon \cite{Israel:1967za}.
In some cases, static black holes have an event horizon, and in addition the curvature invariants (e.g., $R$, $R_{\mu\nu}R^{\mu\nu}$, $R_{\kappa\lambda\mu\nu}R^{\kappa\lambda\mu\nu}$) take finite values over the whole range of the radial coordinate $r$. Such black holes have been conventionally called "regular" black holes, although this is an abuse of language (they are more accurately non-singular black holes). In general, the proper definition of a regular black hole requires the study of: (i) the divergence of the curvature invariants, and (ii) the incompleteness of the geodesics (see \cite{Lan:2023cvz} for further details).

Thus, a "regular" black hole represents a solution of the Einstein field equations without singularity in the centre. According to the Penrose singularity theorem~\cite{bib:penrose}, the gravitational collapse of the matter cloud always leads to the singularity formation if the strong energy condition is held. 
Note that Penrose's theorem can be circumvented if, for example, the strong energy condition is violated near the center of a black hole.
The existence of a singularity usually means that (at least) one of their invariants (the Kretschmann scalar $K=R_{iklm}R^{iklm}$, for instance)
%
is divergent in the limit $r\to 0$. 

Bardeen was the first who constructed the black hole solution with regular center~\cite{bib:bardeen}, i.e. the black hole solution with regular Kretschmann scalar in the whole spacetime. Later, it was understood that the Bardeen solution is supported by nonlinear electrodynamics~\cite{bib:bardeen_non}. Today, we have strong experimental evidence of a black hole existence through gravitational waves detection by LIGO and VIRGO groups~\cite{bib:ligo1, bib:ligo2} and black hole images in the galaxies M87~\cite{bib:ehtm871, bib:ehtm872} and Milky Way~\cite{bib:ehtcrg1, bib:ehtcrg2}. Meanwhile, the existence of a singularity is an indication that general relativity is not fully applicable in this high-density region. Thus, black hole models with non-singular center appeal to the attention of the scientific community~\cite{bib:bardeen, bib:hay, bib:dym1, bib:dym2, bib:charged_regular, bib:rinkon, bib:zaslav_regular, bib:dynamix, bib:charged2, bib:thermo_regular, bib:maharaj_dym, bib:maharaj_rotating, bib:vaidya_nonsingular, bib:aizrek, bib:joshi_regular, bib:thermo_hay, bib:bsw_regular, bib:vermax2, bib:ovalle_regular, bib:ghosh2023, bib:baptista2024,Rincon:2020cos}. For detailed reviews, see~\cite{bib:ansoldi, bib:regular2023}.

Subsequently, several authors investigated novel static regular black hole metrics, one of the most notable being the one introduced by Hayward \cite{bib:hay}. 
Thus, the Hayward black hole metric is 
  i) static, 
 ii) without electric charge, and
iii) without a cosmological constant, 
in a spherically symmetric spacetime. 
Furthermore, the Hayward solution has an intrinsic parameter, $l$, which encodes the deviation concerning the Schwarzschild black hole solution. Note also that this metric becomes a de-Sitter spacetime at the center of the black hole, which is the reason why there is no singularity (at $r=0$) and, in addition, the solution is asymptotically flat for large values of $r$.
Although the Hayward black hole metric was first obtained from the modified Einstein equations, such a result in the context of general relativity can be found equivalently by taking advantage of concrete nonlinear electrodynamics \cite{bib:bardeen_non,Ayon-Beato:1999kuh}, in which case the auxiliary parameter can be reinterpreted in terms of a magnetic charge.
Regular black holes have also been studied from a theoretical perspective in some papers, for example \cite{Bargueno:2021fus,Melgarejo:2020mso,Bargueno:2020ais,Morales-Duran:2016jqt} and references therein.

At this point becomes convenient to mention that the study of black hole thermodynamics is relevant to several areas of theoretical physics because it mixes several ingredients: spacetime, gravity, and quantum mechanics. Hawking's discovery of black hole radiation, which showed that black holes produce thermal radiation as a result of quantum phenomena near their event horizons, is one of the seminal achievements in this field (see \cite{Hawking:1975vcx} and subsequent citations of this work). This seminal discovery, usually referred to as Hawking radiation, established an undeniable link between black hole physics and thermodynamics, and opened the door to the rich field of black hole thermodynamics research.
Roughly speaking, the black hole thermodynamic properties include, but are not limited, to the study of their entropy, temperature, and energy, in order to make progress in connection with gravity, quantum field theory, and statistical mechanics. Nowadays it is well-known that the microscopic origin of black hole entropy links it to the quantum states of fields near the horizon \cite{Bekenstein:1973ur}. This connection has led to significant advances in our understanding of the holographic principle and the AdS/CFT correspondence, providing powerful insights into the quantum nature of spacetime \cite{Maldacena:1997re}.
The thermodynamics of black holes have significant implications for astrophysics and cosmology; for example, the thermodynamic behaviors that govern processes such as black hole accretion, evaporation, and interactions with their surroundings are central to shaping the properties of galaxies, galactic nuclei, and the overall structure of the cosmos \cite{Narayan:2013gca}.

Recently, a new method for solving the Einstein field equations has been proposed~\cite{bib:gd1,bib:gd2}. It has been proven that it is possible to solve the Einstein field equations for the matter, which energy-momentum tensor is given by
\begin{equation}
\hat{T}_{ik}=T_{ik}+\Theta_{ik}.
\end{equation}
Firstly, one should find the solution for the matter source $T_{ik}$ and then for $\Theta_{ik}$ separately, then, by straightforward superposition of two solutions, one can obtain the complete solution for the source $\hat{T}_{ik}$. In particular, it means that if we know the solution of the Einstein field equations with matter source $T_{ik}$ then one can consider the more realistic energy-momentum tensor by adding a new matter source $\Theta_{ik}$ which causes the deformation of the geometry. If this matter source deforms only $g_{11}$ component of the metric tensor, then this method is called minimal geometrical deformation [MGD]~\cite{bib:mgd1, bib:mgd2}. If this matter source deforms both $g_{00}$ and $g_{11}$ components, then it is called an extended gravitational decoupling [EGD]~\cite{bib:gd1, bib:gd2, bib:gd3, bib:gd4}. The known solution of the Einstein field equations with energy-momentum tensor $T_{ik}$ is called a seed metric. 

The Einstein field equations are non-linear and MGD and EDG methods are powerful tools which can help to obtain new solutions with more realistic matter. Moreover, it has been shown~\cite{bib:bh1, bib:bh2}, that gravitational decoupling can lead to a hairy black hole solution. The gravitational decoupling has been applied to several well-known solutions of the Einstein field equations~\cite{bib:vermax, bib:rotating, bib:cosmology, bib:varmholes, bib:kiselev}.
To be more precise, both, the minimal geometric deformation and the extended gravitational decoupling have been significantly used in the last years in the context of black holes, relativistic stars, and cosmological models (see for instance \cite{Maurya:2019sfm,Estrada:2018zbh,Estrada:2018zbh,Cavalcanti:2016mbe,Casadio:2017sze,daRocha:2020jdj,Gabbanelli:2018bhs,Panotopoulos:2018law,Rincon:2019jal,Rincon:2020izv,Tello-Ortiz:2023yxz,Tello-Ortiz:2020nuc,Singh:2019ktp}
 and references therein).

The purpose of this work is twofold: first, we apply gravitational decoupling to regular Hayward and Bardeen black holes to falsify whether an additional matter source can lead to a regular solution or whether it breaks the regularity in the center, and second, we study the corresponding black hole thermodynamics for each case, for a concrete range of the models' parameters.


This work is organized as follows: after this short introduction, in section \eqref{Gra-Dec} we briefly describe the gravitational decoupling method. Subsequently, in sec. \eqref{Hayward} and \eqref{Bardeen} new extensions of Hayward and Bardeen regular black holes are obtained. We show that an extra source might lead to both singular and regular solutions. Some thermodynamic properties of newly obtained solutions are discussed in sec. \eqref{Termo}. Concluding remarks are contained in the last section.

Throughout the paper, the geometrized system of units $G=c=1$ and signature $-+++$ will be used.

\section{Gravitational decoupling and Hairy Schwarzschild black hole}\label{Gra-Dec}

In this section, we briefly describe the gravitational decoupling. 

Gravitational decoupling states that, under some conditions,  one can solve the Einstein field equations with the matter source
\begin{equation}
\tilde{T}_{ik}=T_{ik}+\Theta_{ik} \,,
\end{equation}
where $T_{ik}$ represents the energy-momentum tensor of a system for     which the Einstein field equations are
\begin{equation} \label{eq:thefirst}
G_{ik}=8\pi T_{ik} \,.
\end{equation}
The solution of the equations \eqref{eq:thefirst} is supposed to be known. This solution can be a well-known one of the Einstein field equations. For example, like in our case, it can be Hayward or Bardeen solutions.  It represents the seed solution.
Then $\Theta_{ik}$ represents an extra matter source which causes additional geometrical deformations.
This matter source is obeyed the Einstein field equations, which are given by
\begin{equation} \label{eq:thesecond}
\bar{G}_{ik}=\alpha \Theta_{ik} \,,
\end{equation}
where $\alpha$ is a coupling constant and $\bar{G}_{ik}$ is the Einstein tensor of deformed metric only.
 The gravitational decoupling states that despite of non-linear nature of the Einstein equations, a straightforward superposition  of these two solutions
\eqref{eq:thefirst} and \eqref{eq:thesecond}
\begin{equation}
\tilde{G}_{ik}\equiv G_{ik}+\bar{G}_{ik}=8\pi T_{ik}+\alpha
\Theta_{ik}\equiv \tilde{T}_{ik} \,,
\end{equation}
is, under some conditions,  also the solution of the Einstein field equations.

Let's consider the Einstein field equations
\begin{equation} \label{eq:ex1}
G_{ik}=R_{ik}-\frac{1}{2}g_{ik}R=8\pi T_{ik} \,.
\end{equation}
Let the solution of \eqref{eq:ex1} be a static spherically-symmetric spacetime of the form
\begin{equation} \label{eq:seed}
ds^2=-e^{\nu(r)}dt^2+e^{\lambda(r)}dr^2+r^2 d\Omega^2 \,.
\end{equation}
Here $d\Omega^2=d\theta^2+\sin^2\theta d\varphi^2$ is the metric on unit two-sphere, $\nu(r)$ and $\lambda(r)$ are function of $r$ coordinate and they are supposed to be known.
The metric \eqref{eq:seed} is called as the seed metric.

Now, we seek the geometrical deformation of \eqref{eq:seed} by introducing two new functions $\xi=\xi(r)$ and $\eta=\eta(r)$ by:
\begin{eqnarray} \label{eq:deform}
e^{\nu(r)} &\rightarrow & e^{\nu(r)+\alpha \xi(r)},\nonumber \\
e^{\lambda(r)} &\rightarrow & e^{\lambda(r)}+\alpha \eta(r),
\end{eqnarray}
here $\alpha$ is a coupling constant. Functions $\xi$ and $\eta$ are associated with geometrical deformations of $g_{00}$ and $g_{11}$ of the metric \eqref{eq:seed} respectively. These deformations are caused by new matter source $\Theta_{ik}$. If one puts $\xi(r)\equiv 0$ then the only $g_{11}$ component is deformed, leaving $g_{00}$ unchanged-this is the minimal geometrical deformation. It has some drawbacks, for example, by using this method it is hard to obtain a stable black hole with a properly defined event horizon~\cite{bib:bh2}. If we deform both $g_{00}$ and $g_{11}$ components then this is an extended gravitational decoupling. 

Substituting \eqref{eq:deform} into \eqref{eq:seed}, one obtains:
\begin{equation} \label{eq:noseed}
ds^2=-e^{\nu+\alpha \xi}dt^2+\left(e^{\lambda}+\alpha \eta \right)
dr^2+r^2 d\Omega^2 \,.
\end{equation}
The Einstein equations for \eqref{eq:noseed}
\begin{equation}
\tilde{G}_{ik}= 8 \pi \tilde{T}_{ik}=8\pi (T_{ik}+\Theta_{ik} ) \,,
\end{equation}
are
\begin{equation} \label{eq:einstein}
    \begin{split}
&8\pi (T^0_0+\Theta^0_0)=-\frac{1}{r^2}+e^{-\beta}\left( \frac{1}{r^2}-\frac{\beta'}{r}\right), \\\\
&8\pi (T^1_1+\Theta^1_1)=-\frac{1}{r^2}+e^{-\beta}\left(\frac{1}{r^2}+\frac{\nu'+\alpha \xi'}{r}\right), \\\\
&8\pi (T^2_2+\Theta^2_2)=\frac{1}{4}e^{-\beta}\left[2(\nu''+\alpha \xi'')+(\nu'+\alpha \xi')^2 \right. \\\\ & \left. -\beta'(\nu'+\alpha \xi') 
+2\frac{\nu'+\alpha \xi'-\beta'}{r} \right], \\\\
&e^{\beta}\equiv e^{\lambda}+\alpha \eta,
    \end{split}
\end{equation}
here, the prime denotes partial derivative with respect to radial coordinate $r$ and $8\pi T^2_2+\Theta^2_2=8 \pi T^3_3+\Theta^3_3$ due to spherical symmetry.

From \eqref{eq:einstein} one can define the effective energy density
$\tilde{\rho}$, effective radial $\tilde{P}_r$ and effective tangential $\tilde{P}_t$ pressures as
\begin{equation}  \label{eq:effective}
    \begin{split}
\tilde{\rho}&=-(T^0_0+\Theta^0_0), \\
\tilde{P}_r&=T^1_1+\Theta^1_1, \\
\tilde{P}_t&=T^2_2+\Theta^2_2 .
    \end{split}
\end{equation}
From \eqref{eq:effective} one can introduce the anisotropy parameter $\Pi$ as
\begin{equation} \label{eq:anisotropy}
\Pi=\tilde{P}_t-\tilde{P}_r \,,
\end{equation}
if $\Pi\neq 0$ then it indicates the anisotropic behaviour of fluid $\tilde{T}_{ik}$.

The equations \eqref{eq:einstein} can be decoupled into two
parts\footnote{One should remember that it always works for
    $T_{ik}\equiv 0$ i.e. the vacuum solution and for special cases of
    $T_{ik}$ if one opts for Bianchi identities
    $\nabla_iT^{ik}=\nabla_i\Theta^{ik}=0$ with respect to the metric
    \eqref{eq:noseed} otherwise there an energy exchange i.e.
    $\nabla_i\tilde{T}^{ik}=0\rightarrow
    \nabla_iT^{ik}=-\nabla_i\Theta^{ik}\neq 0$.}:
the Einstein equations corresponding to the seed solution \eqref{eq:seed} and the one corresponding to the geometrical deformations.
If we consider the
vacuum solution i.e. $T_{ik}\equiv 0$ - Schwarzschild solution then, by solving the Einstein field equations which correspond the geometrical deformations, one obtains the hairy Schwarzschild  solution~\cite{bib:bh1}
\begin{equation} \label{eq:seedsch}
    \begin{split}
        ds^2&=-\left(1-\frac{2\mathcal{M}}{r}+\alpha e^{-\frac{r}{\mathcal{M}-\frac{\alpha
L}{2}}}\right) dt^2 \\\\ & +\left(1-\frac{2\mathcal{M}}{r}+\alpha e^{-\frac{r}{\mathcal{M}-\frac{\alpha
L}{2}}}\right) ^{-1}dr^2+r^2d\Omega^2 \,,
    \end{split}
\end{equation}
where $\alpha$ is the coupling constant, $L$ is a new parameter with length dimension and associated with a primary hair of a black hole. $\mathcal{M}$ is the mass of the black hole, which relates to the Schwarzschild mass $M$ by the relation
\begin{equation}
\mathcal{M}=M+\frac{\alpha L}{2} \,,
\end{equation}
the impact
of $\alpha$ and $L$ on the geodesic motion, gravitational lensing,
energy extraction   and the thermodynamics has been considered
in~\cite{bib:geod, bib:lens, bib:energy, bib:thermo}.
The effective pressure and energy density for the metric \eqref{eq:seedsch} are given by
\begin{equation} \label{eq:press-den}
    \begin{split}
        \tilde{P}_r&=-\tilde{\rho}=-\frac{r-M}{8\pi Mr^2}\,\alpha e^{-\frac{r}{M}}, \\\\
\tilde{P}_t&=\frac{r-2M}{16 \pi M^2r}\,\alpha e^{-\frac{r}{M}},
    \end{split}
\end{equation}
One can see that the matter distribution is anisotropic. The average pressure $\bar{P}$ is given by

\begin{equation}
\bar{P}=\frac{1}{3}\left(\tilde{P}_r+2\tilde{P}_t\right )=\frac{r^2+M^2-3Mr}{8\pi M^2r^2}\alpha e^{-\frac{r}{M}}.
\end{equation}
One can introduce the effective parameter of the equation of state $w_{eff}$ by
\begin{equation}
\frac{\bar{P}}{\tilde{\rho}}=w_{eff}=\frac{r^2+M^2-3Mr}{M(r-M)}.
\end{equation}
Note, that the parameter $w_{eff}$ is $r$-depended, but doesn't depend on both a coupling constant $\alpha$ and a primary hair $L$.
The relative pressure anisotropy is given by
\begin{equation}
\Delta=\frac{\Pi}{\bar{P}}=\frac{1+w_{eff}}{2w_{eff}}.
\end{equation}
This expression shows anisotropic behavior of $\Theta_{ik}$ because $w_{eff}$ is not a constant  and can take $w_{eff}=-1$ only at some radius $r$.

 \section{Hairy Hayward regular black hole}\label{Hayward}

In this section we consider Hayward regular black hole as a seed spacetime and consider if one can keep a regular center by applying the gravitational decoupling. Considering the process of formation and evaporation of a regular black hole, Hayward~\cite{bib:hay} offered a minimal model of a regular black hole
 \begin{equation} \label{eq:hay}
ds^2 = - \left( 1-\frac{2 M r^{2}}{r^{3}+2
l^{2} M} \right) dt^2 +\left( 1-\frac{2 M r^{2}}{r^{3}+2
l^{2} M} \right)^{-1} dr^2 + r^2 d \Omega^2 \,,
\end{equation}
here $M$ is a mass of a black hole and $l$ is a positive regularization constant. This solution \eqref{eq:hay} reduces to  Schwarzschild black hole by setting $l=0$ and flat spacetime if $M=0$.
The spacetime~\eqref{eq:hay} behaves at the centre as:
 \begin{equation} \label{eq:zerohay}
  f(r) \simeq 1 - \frac{r^2}{l^2 } \,, r
\rightarrow 0 \,.
 \end{equation}
Which is similar to de Sitter spacetime.
On the other hand, in infinity it behaves as a Schwarzschild solution
 \begin{equation} \label{eq:infhay}
f(r) \simeq 1 - \frac{2 M}{r} \,, r
\rightarrow \infty \,,
\end{equation}
The Einstein tensor components for \eqref{eq:hay} are given by
 \begin{eqnarray} \label{eq:einsteinhay}
  &&G_{0}^{0} = G_{1}^{1} = -\frac{12 M^{2} l^{2}}{\left(2 l^{2}
  M+r^{3}\right)^{2}}\, \\
  &&G_{2}^{2} = G_{3}^{3} = -\frac{24 \left(l^{2} M-r^{3}\right) M^{2}
   l^{2}}{\left(2 l^{2} M+r^{3}\right)^{3}}
 \end{eqnarray}
The solution \eqref{eq:hay} is supported by the energy-momentum tensor 
\begin{equation} \label{eq:emthay}
    \begin{split}
        \rho&=-T^0_0=\frac{1}{8\pi}\frac{12M^2l^2}{\left(2Ml^2+r^3\right)^2}, \\
P_r&=T^1_1=-\frac{1}{8\pi}\frac{12M^2l^2}{\left(2Ml^2+r^3\right)^2},\\
P_t&=T^2_2=T^3_3=\frac{1}{8\pi}\frac{24M^2l^2(r^3-Ml^2)}{\left(2Ml^2+r^3\right)^3},
    \end{split}
\end{equation}
here $\rho$ is the energy density, $P_r$ and $P_t$ are radial and tangent pressure. As one can see from \eqref{eq:emthay}, the matter and energy are obeyed the following equations of the state
\begin{equation}\label{eq:statehay}
    \begin{split}
        P_r&=-\rho,
\\
P_t&=\frac{2(r^3-Ml^2)}{r^3+2Ml^2}\rho .
    \end{split}
\end{equation}
By assuming the  weak energy condition $\rho \geq 0$, the second relation \eqref{eq:statehay} implies that there is the region $0<r<r_c=(Ml^2)^{\frac{1}{3}}$ where tangent pressure is negative and it is positive outside this region $r>r_c$. The energy-momentum tensor \eqref{eq:emthay} is anisotropic, and the anisotropy parameter $\Pi$ is
\begin{equation}
\Pi=P_r-P_t=\frac{3r^3}{r^3+2Ml^2}P_r \,,
\end{equation}
which is zero only at $r=0$.

Now, we apply the gravitational decoupling and write the energy momentum tensor in the form
\begin{equation}
\hat{T}_{ik}=8\pi(T_{ik}+\alpha \Theta_{ik}),
\end{equation}
where $T_{ik}$ is the energy-momentum tensor of Hayward spacetime \eqref{eq:emthay}. By solving the Einstein field equation 
$ G_{ik}=8\pi T_{ik} $, 
one obtains Hayward regular spacetime \eqref{eq:hay}.
 $ \alpha \Theta^0_0=\alpha\Theta^1_1=-\tilde{\rho} $, $ \alpha \Theta^2_2=\alpha \Theta^3_3=\tilde{P}_t $
 is the energy-momentum tensor \eqref{eq:press-den}.
 By solving the Einstein field equation 
$ \tilde G_{ik}=8\pi \alpha \Theta_{ik} $ for the spherically-symmetric spacetime
\begin{equation}
ds^2=-f(r)dt^2+f^{-1}(r)dr^2+r^2d\Omega^2,
\end{equation}
one obtains 
\begin{equation}
f(r)=1-\frac{\alpha L}{r}+\alpha e^{-\frac{r}{M}}.
\end{equation}
Now, by solving the  Einstein field equation $\hat{G}_{ik}=\tilde{G}_{ik}+G_{ik}=\hat{T}_{ik}$ and defining the integration constant as $-2M-\alpha L$, we obtain hairy Hayward spacetime in the form 
  \begin{eqnarray} \label{eq:hairyhaybad}
  ds^2 &= & - \left( 1-\frac{2 M \,r^{2}}{2 l^{2} M+r^{3}}-\frac{\alpha L}{r}+\alpha \exp\left(
 \frac{-r}{M} \right)
\right) dt^2\nonumber \\ 
&& +\left(1-\frac{2 M \,r^{2}}{2 l^{2} M+r^{3}}-\frac{\alpha L}{r}+\alpha  \exp\left(
 \frac{-r}{M} \right)
\right)^{-1} dr^2 + r^2 d \Omega^2.
    \end{eqnarray}
If we put $l=0$ one obtains the hairy Schwarzschild solution \eqref{eq:seedsch}. If $ \alpha =0 $, we obtain Hayward regular black hole.

However, the solution~\eqref{eq:hairyhaybad} is not regular black hole anymore. The Kretschmann scalar
 \begin{equation}
  \lim_{r\to 0} K\equiv R_{ i k l m } R^{ i k l m }\,  = \infty \,,
\end{equation}
is divergent in the centre. 

Thus, if one opts for the regular black hole solution by gravitational decoupling, the energy-momentum tensor \eqref{eq:press-den} does not suit this purpose.

In order to obtain a regular black hole by using the gravitational decoupling, one shall proceed in two different ways:
\begin{enumerate}
\item If one introduces the geometrical deformation to Hayward solution by assuming
\begin{equation}
1-\frac{2Mr^2}{r^3+2Ml^2}\rightarrow 1-\frac{2Mr^3}{r^3+2Ml^2}+\alpha g(r),
\end{equation}
then one can construct a regular black hole by proper choice of the function $g(r)$. For example, if one chooses 
\begin{equation}
g(r)=e^{-\frac{r^3+2Ml^2}{r^2M}},
\end{equation}
then the total spacetime
\begin{equation} \label{eq:haymet2}
    \begin{split}
       &  ds^2 = - \left( 1-\frac{2 M \,r^{2}}{2 l^{2} M+r^{3}}+\alpha
  \exp \left({\frac{-2 l^{2} M-r^{3}}{r^{2}
  M}} \right) \right) dt^2   \\\\
 & +\left( 1-\frac{2 M \,r^{2}}{2 l^{2} M+r^{3}}+\alpha
  \exp \left({\frac{-2 l^{2} M-r^{3}}{r^{2}
  M}} \right) \right)^{-1} dr^2 + r^2 d \Omega^2,
    \end{split}
\end{equation}
is regular black hole. One should note, that despite this solution can be obtained by gravitational decoupling it is not hairy Hayward solution because the limit $l\rightarrow 0$ leads to spacetime \eqref{eq:seedsch} with zero valued primary hair $L$.
The Kretschmann scalar is finite in the centre
 \begin{equation} \label{eq:krbad}
  \lim_{r\to 0} R_{ i k l m } R^{ i k l m }\,  = \frac{24}{l^4}\,,
\end{equation}
The Kretschmann scalar does not possess any other zeros of the denominator:

\begin{equation}
K \sim \frac{1}{r^{12} M_{sh}^{4} \left(2 l^{2} M+r^{3}\right)
^{6}} \,,
 \end{equation}
and the only bad point is $r=0$ but as one can see from \eqref{eq:krbad}
 the solution is regular in this limit.
Two other curvature scalars i.e., the squared Ricci tensor
  \begin{equation}
   \lim_{r\to 0 } R_{ i k} R^{ i k} \,
 = \frac{36}{l^4} \,,  \quad  R_{ i k} R^{ i k} \sim  \frac{1}{r^{12}
    M_{sh}^{4} \left(2 l^{2} M+r^{3}\right)
^{6}} \,,
 \end{equation}
and Ricci scalar 
\begin{equation}
  \lim_{r\to 0} R \,
 = \frac{12}{l^2} \,,  \quad  R \sim  \frac{1}{r^{6} M_{sh}^{2} \left(2
  l^{2} M+r^{3}\right)^{3}}
 \end{equation}
are also finite in the limit $\lim\limits_{r\rightarrow 0}$ and do not 
 have other zeros of the denominator.

This solution has been obtained in~\cite{bib:vermax2}. However, one has a problem with the Tailor series expansion of the spacetime \eqref{eq:haymet2} in the $r\to 0$ and $r\to \infty$ in order to obtain the de Sitter-like solution in the centre and Schwarzschild in the infinity. This fact leads us to the conclusion that this solution should be also changed in order to provide the good Taylor series expansion in the centre and infinity. If the regular solution is not de Sitter-like in the centre then the weak energy condition is violated.
\item We introduce the energy-momentum tensor $\Theta_{ik}$ first and then apply the gravitational decoupling with Hayward spacetime as a seed spacetime.
First of all we assume that spacetime has the following form:
\begin{equation} \label{eq:promhaymet}
\begin{split}
    ds^2=-\left(1-\frac{2Mr^2}{r^3+2Ml^2} + G(r)\right)dt^2+ \\\\ \left(1-\frac{2Mr^2}{r^3+2Ml^2}+G(r)\right)dr^2+r^2d\Omega^2\,.
\end{split}
\end{equation}
The energy-momentum tensor for this spacetime is in the form
\begin{equation} \label{eq:einstein_dim_hay}
\hat{T}_{ik}=8\pi(T_{ik}+\theta_{ik}),
\end{equation}
where $T_{ik}$ is the energy-momentum tensor \eqref{eq:emthay} of the Hayward spacetime. The $\Theta_{ik}$part is related to geometrical deformation $G(r)$ and, following Dymnikova~\cite{bib:dym1}, we assume that this parts corresponds to the anisotropic vacuum.
The vacuum is defined as such a kind of matter which does not allow any preferred reference frame connected with it. As the result, any reference frame is comoving with the vacuum. The solution \eqref{eq:promhaymet} is spherically-symmetric. It means that the energy-momentum tensor, with off-diagonal terms are zero, should have the following form
\begin{equation}
\Theta^0_0=\Theta^1_1,~~ \Theta^2_2=\Theta^3_3.
\end{equation}
The energy-momentum tensor of this form, according to the Petrov algebraic classification has an infinite set of comoving reference frames. So, this energy-momentum tensor can be thought as the energy-momentum tensor of the spherically-symmetric vacuum~\cite{bib:sakharov, bib:gliner}. We assume $\Theta_{ik}$ has the following form
\begin{equation} \label{eq:vacdim}
\rho =-\Theta^0_0= \varepsilon_0 e^{-\frac{r^3}{\alpha L\lambda^2}}\,,
\end{equation}
here $\varepsilon_0$ is the non-zero vacuum density and $\lambda$ is connected to the $\varepsilon_0$ via de Sitter relation (restoring $c$ and $G$)
\begin{equation}
\lambda^2=\frac{3c^4}{8\pi G\varepsilon_0} \,.
\end{equation}
From the fact that $\Theta^0_0=\Theta^1_1$, the Einstein equations give for the metric \eqref{eq:promhaymet} with energy momentum tensor \eqref{eq:einstein_dim_hay} with assumption \eqref{eq:vacdim}
\begin{equation}
\frac{1}{r^2}\left(G'r+G\right)=-\varepsilon_0e^{-\frac{r^3}{\alpha L\lambda^2}} \,.
\end{equation}
The solution of this differential equation is
\begin{equation}
G(r)=-\frac{\alpha L}{r}\left(1-e^{-\frac{r^3}{\alpha L\lambda^2}}\right) \,.
\end{equation}
This solution reminds Dymnikova solution~\cite{bib:dym1}.

Now, by solving the combined  Einstein field equations $\hat{G}_{ik}=\hat{T}_{ik}$, we obtain the metric in the form:
\begin{equation} \label{eq:dymhay}
    \begin{split}
         &ds^2 = -  \left( 1-\frac{2 M \,r^{2}}{2 l^{2}
   M+r^{3}}  -\frac{\alpha  L \left(1-e^{-\frac{r^{3}}{\alpha  L
  \lambda^{2}}}\right)}{r}  \right)  dt^2 \\\\
& +\left( 1-\frac{2 M \,r^{2}}{2 l^{2}
   M+r^{3}}-\frac{\alpha  L \left(1-e^{-\frac{r^{3}}{\alpha  L
  \lambda^{2}}}\right)}{r}  \right) ^{-1} dr^2 + r^2 d \Omega^2.
    \end{split}
\end{equation}
Here, $L$ is a parameter which has a length dimension. It is similar  to hairy Schwarzschild solution and can be associated with a primary hair.
Now, if we look at the Taylor series expansion of \eqref{eq:dymhay} in the centre
 \begin{equation}
  f(r) \simeq 1+\left(-\frac{1}{l^{2}}-\frac{1}{\lambda^{2}}\right) r^{2} \,, r
\rightarrow 0 \,,
 \end{equation}
which is similar to the de Sitter solution.
In infinity one has
 \begin{equation}
f(r) \simeq 1+\frac{-\alpha  L-2 M}{r} \,, r
\rightarrow \infty \,,
\end{equation}
Which behaves like Schwarzschild solution.
The Einstein tensor components for the solution \eqref{eq:dymhay} have
 the following form
 \begin{equation} 
 \begin{split}
\hat{G}_0^0 &= \hat{G}_1^1 = \frac{-12 \left(l^{2} M+\frac{r^{3}}{2}\right)^{2}
   e^{-\frac{r^{3}}{\alpha  L \,\lambda^{2}}}-12 \lambda^{2}
 M^{2} l^{2}}{\lambda^{2} \left(2 l^{2} M+r^{3}\right)^{2}} \,, \\\\
           \hat{G}_2^2 &= \hat{G}_3^3 = \frac{-3}{ \lambda^{4} L \alpha  \left(l^{2}
M+\frac{r^{3}}{2}\right)^{3}} \left(  \left(\alpha  L
\lambda^{2}-\frac{3 r^{3}}{2}\right)  \times \right. \\ & \left.  (l^{2} M
 +\frac{r^{3}}{2})^{3}  e^{-\frac{r^{3}}{\alpha  L
 \lambda^{2}}} + \,l^{2} M^{2} \alpha L \,\lambda^{4}
 (l^{2} M -r^{3})  \right)\,.
      \end{split}
  \end{equation}

The solution \eqref{eq:dymhay} is the composition of Hayward
 solution~\eqref{eq:hay} and Dymnikova solution~\cite{bib:dym1}.
The Kretschmann scalar is regular in the centre
 \begin{equation}
  \lim_{r\to 0} R_{ i k l m } R^{ i k l m }\,  =
  \frac{24}{\lambda^{4}}+\frac{48}{\lambda^{2} l^{2}}+\frac{24}{l^{4}} \,,
\end{equation}
The denominator of Kretschmann scalar is given by

\begin{equation}
K \sim \frac{1}{L^{2} r^{6} \alpha^{2} \left(2 l^{2}
  M+r^{3}\right)^{6} \lambda^{8} }
 \end{equation}
which has only one 'bad' point $r=0$ but the limit above shows that at
 $r=0$ the Kretschmann scalar is regular.
The squared Ricci  tensor at the limit $r\rightarrow 0$ and its the
 denominator has the form:
  \begin{equation}
   \lim_{r\to 0 } R_{ i k} R^{ i k} \,
 = \frac{36}{\lambda^{4}}+\frac{72}{\lambda^{2} l^{2}}+\frac{36}{l^{4}}
   \,,  \quad  R_{ i k} R^{ i k} \sim  \frac{1}{2 \lambda^{8} \left(2
   l^{2} M+r^{3}\right)^{6} \alpha^{2} L^{2}   }
 \end{equation}
And the same for Ricci scalar
\begin{equation}
  \lim_{r\to 0} R \,
 = \frac{12}{\lambda^{2}}+\frac{12}{l^{2}} \,,  \quad  R \sim
  \frac{1}{\left(2 l^{2} M+r^{3}\right)^{3} \lambda^{4}
   \alpha  L  }
 \end{equation}
The energy-momentum tensor of the whole spacetime \eqref{eq:dymhay} is
\begin{equation} \label{eq:emtdymhay}
    \begin{split}
         \hat{\rho} & =-\hat{P}_r \rightarrow -\hat{T}^0_0=-\hat{T}^1_1 =\varepsilon_0 e^{-\frac{r^3}{\alpha L \lambda^2}}\,+ \\ \\ &  + \frac{1}{8\pi}\frac{12M^2l^2}{\left(2Ml^2+r^3\right)^2}\equiv 
-T^0_0-\Theta^0_0, \\\\ 
\hat{P}_t & =\hat{T}^2_2=\hat{T}^3_3 =\frac{1}{8\pi}\frac{24M^2l^2(r^3-Ml^2)}{\left(2Ml^2+r^3\right)^3} + \\\\
&+\frac{1}{8\pi}\frac{3r^3-2\alpha L \lambda^2}{2\alpha L \lambda^2}\varepsilon_0 e^{-\frac{r^3}{\alpha L \lambda^2}}=T^2_2+\Theta^2_2.
    \end{split}
\end{equation}
Here we have introduced effective energy density $\hat{\rho}$, radial   $\hat{P}_r$ and tangent $\hat{P}_t$ pressure
\begin{equation}
    \begin{split}
        \hat{\rho}&=\rho^{hayward}+\rho^{dymnikova}, \\
\hat{P}_r&=P_r^{hayward}+P_r^{dymnikova},\\
\hat{P}_t&=P_t^{hayward}+P_t^{dymnikova}.
    \end{split}
\end{equation}
Therefore we have found out the gravitational decoupling of two matter field, i.e. one is interpreted with anisotropic vacuum and other matter source is the usual matter and energy distribution of the Hayward spacetime.
\end{enumerate}

 \section{Hairy Bardeen regular black hole}\label{Bardeen}

In this section, we apply the gravitational decoupling to Bardeen black hole. We will follow the same steps which are described in the previous section. Bardeen was one of the first who offered the black hole solution with regular centre~\cite{bib:bardeen}. The Bardeen regular  black hole line  element is given by
 \begin{equation} \label{eq:bardeen}
ds^2 = - \left( 1-\frac{2 M \,r^{2}}{\left(E^{2}+r^{2}\right)
^{\frac{3}{2}}} \right) dt^2 +\left(1-\frac{2 M \,r^{2}}{\left
(E^{2}+r^{2}\right)^{\frac{3}{2}}}  \right)^{-1} dr^2 + r^2 d \Omega^2 \,,
 \end{equation}
$M$ is the mass of a black hole and $e$ is magnetic monopole charge. In the case $e=0$ the Bardeen solution \eqref{eq:bardeen} reduces to Schwarzschild black hole solution.
In the center, the solution \eqref{eq:bardeen} behaves like
 \begin{equation}
f(r) \simeq 1 - \frac{2 M}{E^3} r^2  \,, r
\rightarrow 0 \,,
\end{equation}
which is de Sitter-like behavior.
In the limit $r\rightarrow \infty$ the solution \eqref{eq:bardeen}
 behaves like Schwarzschild solution:
\begin{equation}
f(r) \simeq 1 - \frac{2 M}{r} \,, r
\rightarrow \infty \,,
\end{equation}
The Einstein tensor components for the solution \eqref{eq:bardeen} are given by
 \begin{equation} \label{eq:einsteinbardeen}
    \begin{split}
        G_{0}^{0}& = G_{1}^{1} = -\frac{6 M \,E^{2}}{\left(E^{2}+r^{2}\right)
  ^{\frac{5}{2}}}, \\\\
G_{2}^{2} &= G_{3}^{3} = -\frac{6 E^{2} \left(E^{2}-\frac{3
  r^{2}}{2}\right) M}{\left(E^{2}+r^{2}\right)^{\frac{7}{2}}}.
    \end{split}
\end{equation}
From the Einstein tensor \eqref{eq:einsteinbardeen}, one can obtain the energy-momentum tensor of the form
\begin{eqnarray} \label{eq:bardeennew}
T_0^0&=&T^1_1=-\frac{1}{4\pi}\frac{6 M \,E^{2}}{\left(E^{2}+r^{2}\right)
  ^{\frac{5}{2}}},\nonumber \\
T^2_2&=&T^3_3=-\frac{1}{4\pi}\frac{6 E^{2} \left(E^{2}-\frac{3
  r^{2}}{2}\right) M}{\left(E^{2}+r^{2}\right)^{\frac{7}{2}}}.
 \end{eqnarray}

If we look for the Einstein equation solution with right-hand side of the form:
\begin{equation}
\hat{T}_{ik}=T_{ik}+\alpha \Theta_{ik},
\end{equation}
where $T_{ik}$ is the energy-momentum tensor corresponding to the Bardeen solution \eqref{eq:bardeennew} and $\alpha \Theta_{ik}$ is an anisotropic fluid \eqref{eq:press-den} then we obtain the solution of the form:

  \begin{eqnarray} \label{eq:bardeenbad}
  ds^2 &= & - \left(1-\frac{2 M \,r^{2}}{\left(E^{2}+r^{2}\right)
  ^{\frac{3}{2}}}-\frac{\alpha L}{r}+\alpha e^{-\frac{r}{M}}  \right) dt^2\nonumber
  \\ && +\left(1-\frac{2 M \,r^{2}}{\left(E^{2}+r^{2}\right)
  ^{\frac{3}{2}}}-\frac{\alpha L}{r}+\alpha e^{-\frac{r}{M}}\right)^{-1} dr^2 + r^2 d
  \Omega^2 \,,
\end{eqnarray}
Like in the Hayward regular black hole case, the extra matter field spoils the regularity condition in the centre and the Kretschmann scalar
\begin{equation}
  \lim_{r\to 0} R_{ i k l m } R^{ i k l m }\,  = \infty \,,
 \end{equation}
is divergent in the centre.
We can choose the deformation function $g(r)$ in the form:
\begin{equation}
g(r)=e^{-\frac{(r^2+E^2)^{\frac{3}{2}}}{r^2M}}.
\end{equation}
Then, if we assume the lapse function is
\begin{equation}
f(r)=1-\frac{2Mr^2}{(r^2+E^2)^{\frac{3}{2}}}+\alpha g(r),
\end{equation}
the deformed Bardeen spacetime becomes
\begin{eqnarray} \label{eq:bardeenhairbad}
  ds^2 &=  & - \left(1-\frac{2 M \,r^{2}}{\left(E^{2}+r^{2}\right)
  ^{\frac{3}{2}}}+\alpha  e^{-\frac{\left(E^{2}+r^{2}\right)
  ^{\frac{3}{2}}}{M \,r^{2}}}  \right) dt^2\nonumber  \\
&&  +\left(1-\frac{2 M \,r^{2}}{\left(E^{2}+r^{2}\right)
  ^{\frac{3}{2}}}+\alpha  e^{-\frac{\left(E^{2}+r^{2}\right)
  ^{\frac{3}{2}}}{M \,r^{2}}}  \right)^{-1} dr^2 + r^2 d
  \Omega^2 \,,
\end{eqnarray}
The Kretschmann scalar is regular in the centre and the denominator has only one peculiar point at $r=0$:
 \begin{equation}
  \lim_{r\to 0} R_{ i k l m } R^{ i k l m }\,  = \frac{96 M^{2}}{E^{6}}\,,
\quad K \sim \frac{1}{ \left(E^{2}+r^{2}\right)^{7} M^{4}
  r^{12}}
 \end{equation}
The squared Ricci tensor  
\begin{equation}
   \lim_{r\to 0 } R_{ i k} R^{ i k} \,
 = \frac{144 M^{2}}{E^{6}} \,,  \quad  R_{ i k} R^{ i k} \sim  \frac{1}{ 2 \left
   (E^{2}+r^{2}\right)^{7} M^{4} r^{12}}
 \end{equation}
and Ricci scalar 
\begin{equation}
  \lim_{r\to 0} R \,
 = \frac{24 M}{E^{3}}\,,  \quad  R \sim  \frac{1}{ \left(E^{2}+r^{2}\right)
  ^{\frac{7}{2}} M^{2} r^{6}}
 \end{equation}
are also regular in the centre. However, in the centre and the infinity this solution does not behave like de Sitter and Schwarzschild's solutions respectively.
Thus, to obtain the solution which has de Sitter-like behavior in the centre and Schwarzschild-like in the infinity, we assume that the energy-momentum tensor $\alpha \Theta_{ik}$ has the same form like in the previous section \eqref{eq:vacdim}. By solving the Einstein field equations, one obtains the hairy Bardeen regular black hole
  \begin{eqnarray} \label{eq:bardeendym}
  ds^2 &=& -  \left( 1-\frac{2 M \,r^{2}}{\left
  (E^{2}+r^{2}\right)^{\frac{3}{2}}}  -\frac{\alpha  L \left
  (1-e^{-\frac{r^{3}}{\alpha  L
  \lambda^{2}}}\right)}{r}  \right)  dt^2\nonumber  \\
&& +\left( 1-\frac{2 M \,r^{2}}{\left(E^{2}+r^{2}\right)
  ^{\frac{3}{2}}} -\frac{\alpha  L \left(1-e^{-\frac{r^{3}}{\alpha  L
  \lambda^{2}}}\right)}{r}  \right) ^{-1} dr^2 + r^2 d \Omega^2 \,,
\end{eqnarray}
This solution behaves like de Sitter solution in the centre 
 \begin{equation}
  f(r) \simeq 1+\left(-\frac{2
  M}{E^{3}}-\frac{1}{\lambda^{2}}\right) r^{2}  \,, r
\rightarrow 0 \,,
 \end{equation}
and like Schwarzschild in the infinity
 \begin{equation}
f(r) \simeq 1+\frac{-\alpha  L-2 M}{r} \,, r
\rightarrow \infty \,,
\end{equation}
The Einstein tensor components for the solution \eqref{eq:bardeendym}
 have the following form
 \begin{equation}
\begin{split}
    G_0^0 &= G_1^1 = \frac{ -3 \left(\left(E^{2}+r^{2}\right)^{\frac{5}{2}}
 e^{-\frac{r^{3}}{\alpha L \,\lambda^{2}}}+2 E^{2} M
\lambda^{2}\right)}{\left(E^{2}+r^{2}\right)^{\frac{5}{2}} \lambda^{2}} \,,\\\\
G_2^2 &= G_3^3 = \frac{-3}{\left(E^{2}+r^{2}\right)^{\frac{7}{2}}
 \lambda^{4} \alpha  L} \left( \left(E^{2}+r^{2}\right)^{\frac{7}{2}}
\times \right. \\\\ & \left.   (\alpha  L
\lambda^{2}-\frac{3 r^{3}}{2})  e^{-\frac{r^{3}}{\alpha  L
\lambda^{2}}}+2 \alpha  L (E^{2}-\frac{3 r^{2}}{2}) \lambda^{4} E^{2}
 M
\right)
\end{split}
\end{equation}
The solution \eqref{eq:bardeendym} is the superposition of two solutions \eqref{eq:bardeen} and Dymnikova one~\cite{bib:dym1}. The sum of Einstein tensor components in Dymnikova and Bardeen spacetimes lead to the Einstein tensor component which corresponds to the spacetime  \eqref{eq:bardeendym}:   
\begin{equation}
    \begin{split}
          G_{0}^{0} &= G_{1}^{1}  = -\frac{6 M \,E^{2}}{\left
  (E^{2}+r^{2}\right)^{\frac{5}{2}}} +\left(-\frac{3  e^{-\frac{r^{3}}{\alpha
  L \lambda^{2}}}}{\lambda^{2}}\right)  \\  \\
                        & =\frac{ -3 \left(\left(E^{2}+r^{2}\right)^{\frac{5}{2}}
 e^{-\frac{r^{3}}{\alpha L \,\lambda^{2}}}+2 E^{2} M
\lambda^{2}\right)}{\left(E^{2}+r^{2}\right)^{\frac{5}{2}} \lambda^{2}}
  \,,\\ \\
   G_{2}^{2} &= G_{3}^{3} = -\frac{6 M \,E^{2} \left
  (E^{2}-\frac{3 r^{2}}{2}\right)}{\left(E^{2}+r^{2}\right)
  ^{\frac{7}{2}}} + \left(-\frac{3 \left(\alpha  L \lambda^{2}-\frac{3
  r^{3}}{2}\right)  e^{-\frac{r^{3}}{\alpha  L
  \lambda^{2}}}}{\lambda^{4} \alpha L}\right)    \\\\
  &= \frac{-3}{\left(E^{2}+r^{2}\right)^{\frac{7}{2}}
 \lambda^{4} \alpha  L} \left( \left(E^{2}+r^{2}\right)^{\frac{7}{2}}
\times \right. \\\\&   \left.  (\alpha  L
\lambda^{2}-\frac{3 r^{3}}{2})  e^{-\frac{r^{3}}{\alpha  L
\lambda^{2}}}+2 \alpha  L (E^{2}-\frac{3 r^{2}}{2}) \lambda^{4} E^{2}
 M \right)
    \end{split}
\end{equation}
The Kretschmann scalar is regular in the centre and its denominator has
only specific point $r=0$:
 \begin{equation}
  \lim_{r\to 0} R_{ i k l m } R^{ i k l m }\,  =  \frac{24 \left(E^{3}+2
  M \lambda^{2}\right)^{2}}{\lambda^{4} E^{6}} \,,
\quad K \sim  \frac{1}{  \left(E^{2}+r^{2}\right)^{7} r^{6} \lambda^{8} \alpha^{2}
  L^{2}}
 \end{equation}
The same for squared Ricci tensor
  \begin{equation}
   \lim_{r\to 0 } R_{ i k} R^{ i k} \, =
  \frac{36 \left(E^{3}+2 M \lambda^{2}\right)
   ^{2}}{\lambda^{4} E^{6}}
   \,,  \quad  R_{ i k} R^{ i k} \sim  \frac{1}{ 2 \left(E^{2}+r^{2}\right)^{7}
   \alpha^{2} L^{2} \lambda^{8}}
 \end{equation}
and Ricci scalar
\begin{equation}
  \lim_{r\to 0} R \,
 =\frac{12 E^{3}+24 M \lambda^{2}}{\lambda^{2} E^{3}} \,,
  \quad  R \sim \frac{1}{ \left
  (E^{2}+r^{2}\right)^{\frac{7}{2}} \lambda^{4} \alpha  L}
 \end{equation}

 \section{Some thermodynamics properties}\label{Termo}
\begin{figure*}[ht]
\centering
\includegraphics[width=0.49\textwidth]{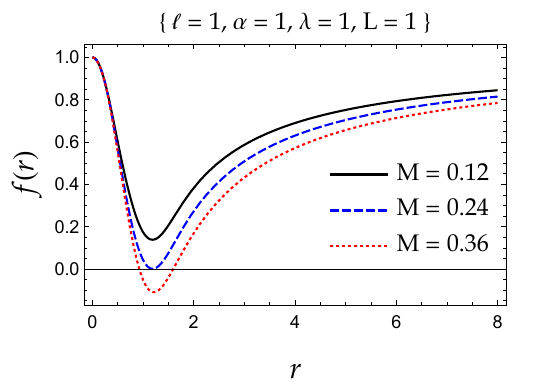}   \
\includegraphics[width=0.49\textwidth]{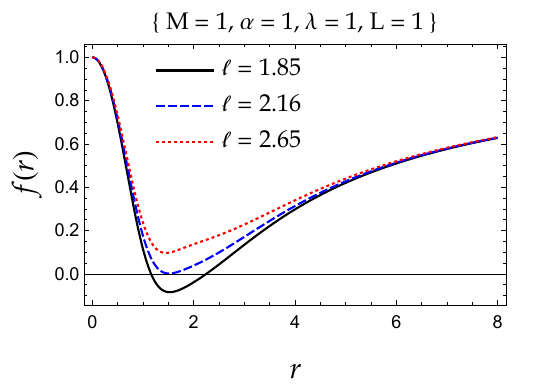}  \
\\
\includegraphics[width=0.49\textwidth]{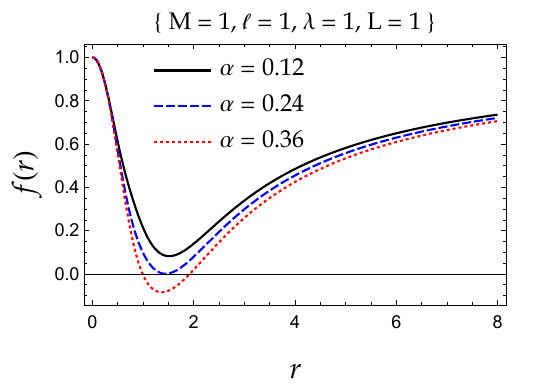}   \
\includegraphics[width=0.49\textwidth]{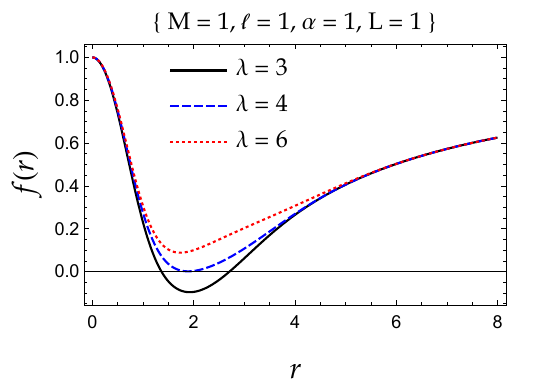}  \
\caption{
Lapse function against the radial coordinate for the Hairy Hayward regular black hole varying the set of parameters 
$\{M, l, \alpha, L, \lambda\}$
%
}
\label{fig:1}
\end{figure*}

\begin{figure*}[ht]
\centering
\includegraphics[width=0.49\textwidth]{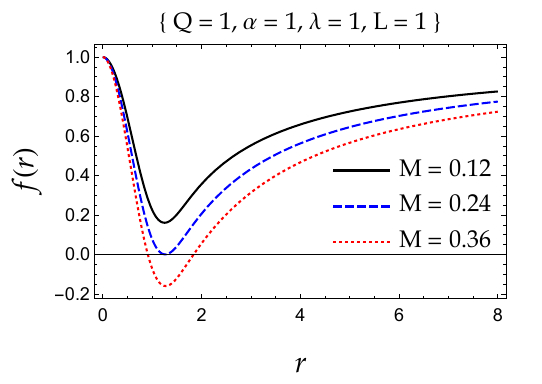}   \
\includegraphics[width=0.49\textwidth]{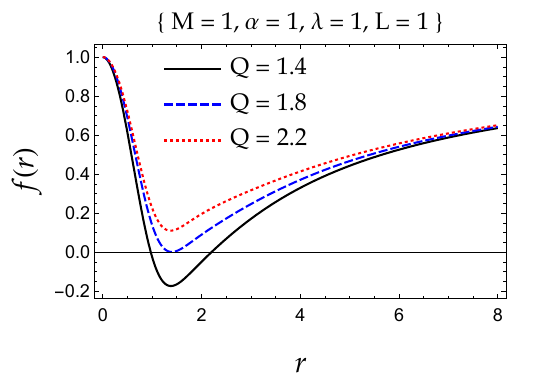}  \
\\
\includegraphics[width=0.49\textwidth]{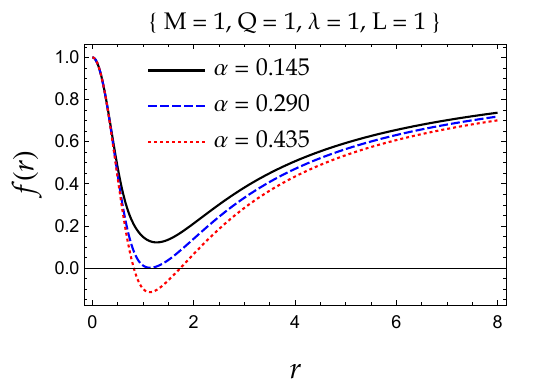}   \
\includegraphics[width=0.49\textwidth]{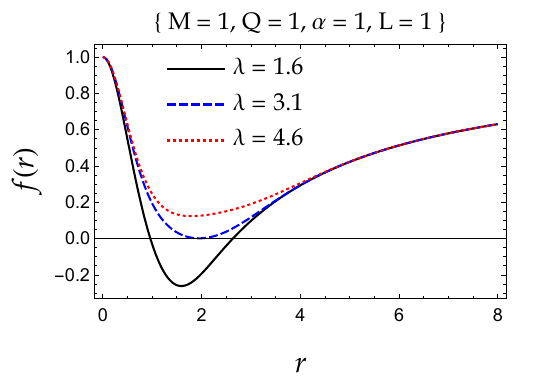}  \
\caption{
Lapse function against the radial coordinate for the Hairy Bardeen regular black hole varying the set of parameters 
$\{M, Q, \alpha, L, \lambda\}$. We have renamed $E$ as $Q$ for practical reasons.
%
}
\label{fig:2}
\end{figure*}

In this section, we will compute some of the basic thermodynamic properties useful to get insights into the classical and quantum nature of a black hole in a four-dimensional space-time. Thus, in what follows, we will investigate: i) the temperature, ii) the entropy, and, iii) the specific heat for the backgrounds used along with this manuscript. 
To do this, we will first compute the event horizon numerically, given the non-trivial form of the lapse function. Once we have $r_H$, we will proceed to calculate the remaining thermodynamic properties.
We consider a metric of four-dimensional static spacetime with the following explicit form 
\begin{align}\label{metric}
    \mathrm{d}s^{2} &=-f(r) \mathrm{d}t^{2} + f(r)^{-1} \mathrm{d}r^{2} + r^{2} \mathrm{d}\Omega^{2}, 
\end{align}
where, as always, the term $\mathrm{d}\Omega^2$ is defined as:
\begin{align}
\mathrm{d}\Omega^2 &= \mathrm{d} \theta^2 + \sin^2 \theta \: \mathrm{d} \phi^2
\label{} \ ,  
\end{align}
where we will consider two concrete forms of the lapse function, the first one, corresponding to the Hairy Hayward regular black hole and the second one corresponding to the Hairy Bardeen regular black hole solution. Lapse functions are plotted in Figs. \eqref{fig:1} and \eqref{fig:2} respectively.
%
\subsection{Hairy Hayward regular black hole}
%
\subsubsection{Event horizon}
Horizons are critical for understanding the structure of a black hole, and in particular for a correct understanding of its thermodynamic properties. In this sense, it is essential to know how the event horizon varies with the black hole mass $M$. 
The event horizon, $r_H$, is the outer root of the lapse function $f(r)$ and is obtained when $f(r_H) = 0$.
Unfortunately, the zero of the lapse function \eqref{eq:dymhay} implies a non-trivial (transcendental) equation for $r$, which cannot be solved analytically.
In the following, we will work with the lapse function \eqref{eq:dymhay}, i.e, 
\begin{align}
    f(r) & = 1-\frac{2 M \,r^{2}}{2 l^{2} M+r^{3}}  - \frac{\alpha L}{r}\bigg(1-e^{-\frac{r^{3}}{\alpha L \lambda^{2}}}\bigg) ,
\end{align}
and the lapse function depends on five parameters, namely $\{M, l, \alpha, L, \lambda \}$. 
We can also redefine the mass function if we notice that the lapse function can also be written as
\begin{align}
f(r) &= 1 - \frac{2 m(r)}{r},
\end{align}
begin $m(r)$ the mass function which takes the form
\begin{align}
m(r) &= \frac{M r^3}{2 l^2 M+r^3}+\frac{1}{2} \alpha  L \bigg(1-e^{-\frac{r^3}{\alpha  \lambda ^2 L}}\bigg).
\end{align}
We solve $f(r_H)=0$ to obtain $r_H$ against $M$ varying the remaining parameters, i.e., $\{l, \alpha, L, \lambda \}$. Notice that the $\alpha$ and $L$ are interchangeable, i.e., the impact of $\alpha$ and $L$ are the same on the event horizon, the reason why we can vary one of them to study the impact of such parameters.
We then compute the event horizon $r_H$ against the black hole mass $M$, varying the parameters $\{ l, \alpha, \lambda \}$ as can be observed in Fig.\eqref{fig:1}. 
\begin{itemize}
    \item From the first row (left), we see that we vary $\alpha$ to fix $\{ l, \lambda, L \}$, and we find that the horizon increases as $\alpha$ increases. In particular, note that the curves look as if they are shifted by a constant value for all the ranges of masses used.
    \item From the first row (middle) we observe that for fixing $\{ \alpha, \lambda, L \}$ we vary $l$ and we found that the horizon increases as $l$ increases.  We observe that for small values of $M$ the effect of $l$ is significant and therefore the event horizon varies considerably, but as $M$ increases the event horizon tends to converge to the same value.
    \item From the first row (right), we observe that when we fix $\{ l, \alpha, L \}$, we vary $\lambda$ and we find that the horizon increases as $\lambda$ increases. In this case, the event horizon is practically independent of $\lambda$, and only some variation appears when $M$ is small. 
\end{itemize}
%
%

%
%

\subsubsection{Temperature}

First, the event horizon of a black hole in four spacetime dimensions is a three-dimensional surface, formed of the null geodesics. A useful concept to be introduced is the so-called Killing horizon, which appears when the spacetime has a symmetry that maps the horizon into itself along the null direction. The importance of such a concept becomes evident due to, in black hole thermodynamics, the temperature $T_H$ of a Killing horizon is identified with the surface gravity $\kappa$ evaluated on a future horizon $H_{+}$ / a past horizon $H_{-}$, and the expression is given according to:

\begin{align}
    T_H &\equiv \frac{\kappa}{2\pi} = 
   \frac{1}{2\pi} \sqrt{-\frac{1}{2} \Bigl(\nabla_{\mu}\chi_{\nu}\Bigl)
    \Bigl(\nabla^{\mu}\chi^{\nu}\Bigl)
    } \ ,
\end{align}
where we have assumed $k_b = 1 = \hbar$ and $\chi^{\nu}$ is a Killing vector (and therefore satisfies that $\nabla_a \chi_b + \nabla_b \chi_a = 0 $, i.e., such equation generates a symmetry of the metric). A more familiar expression, obtained after performing the explicit form of the Killing vector, can be written as
\begin{equation}
T_H =\frac{1}{4\pi}\left|\lim_{r\rightarrow r_{H}}\frac{\partial_{r} g_{tt}}{\sqrt{-g_{tt}g_{rr}}}\right| = \frac{1}{4\pi} \frac{\mathrm{d}f(r)}{\mathrm{d}r}\Bigg |_{r=r_H}
\end{equation}
%
%
The lapse function for the Hairy Hayward regular black hole, albeit analytical, produces a non-trivial form of the Hawking temperature, and the expression is given as follows:
\begin{align}
\begin{split}
    T_H = \frac{1}{4 \pi r_H} 
    \Bigg[
    &\frac{2 M r_H^2 \left(r_H^3 - 4 l^2 M\right)}{\left(2 l^2 M + r_H^3\right)^2} 
    + 
    \Bigg( 
    \frac{2 l^2 M (r_H -\alpha  L)}{r_H \left(2 l^2 M + r_H^3\right)} + 
    \frac{r_H^3 (r_H - \alpha L - 2 M )}{r_H \left(2 l^2 M + r_H^3\right)}
    \Bigg) 
    \\
    &
    \times 
    \Bigg\{
    1 + 3 \ln \left(\frac{\alpha  L \left(2 l^2 M + r_H^3\right)}{r_H^3 (\alpha  L+2 M - r_H)-2 l^2 M (r_H - \alpha  L)}\right)
    \Bigg\}
     + \frac{L \alpha}{r_H}
    \Bigg]
\end{split}
\end{align}

The Hawking temperature can be reduced to the standard case (Schwarzschild black hole solution) by setting $\alpha$, $\lambda$, $L$, and $l$ to zero.

%
%

%
%

\subsubsection{Bekenstein-Hawking entropy}

\begin{figure*}[ht]
\centering
\includegraphics[width=0.32\textwidth]{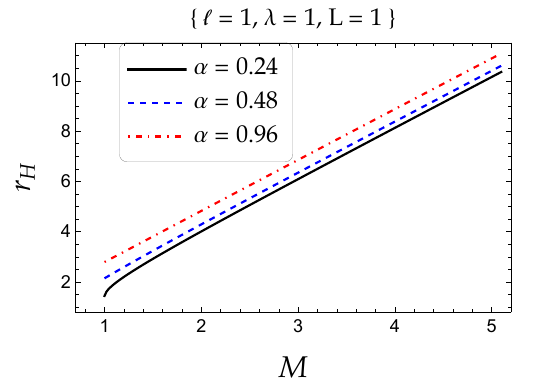}  \
\includegraphics[width=0.32\textwidth]{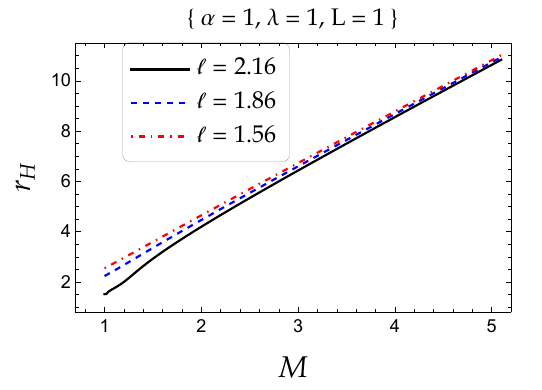}  \
\includegraphics[width=0.32\textwidth]{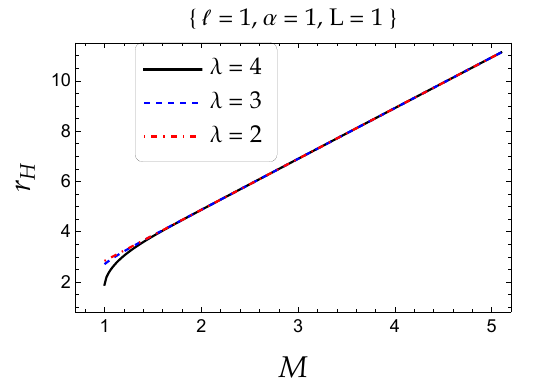}  \
\\
\includegraphics[width=0.32\textwidth]{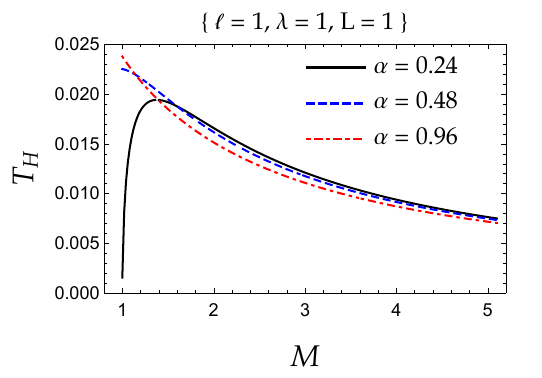}  \
\includegraphics[width=0.32\textwidth]{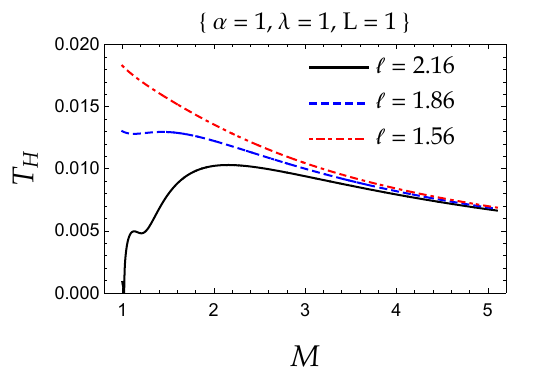}  \
\includegraphics[width=0.32\textwidth]{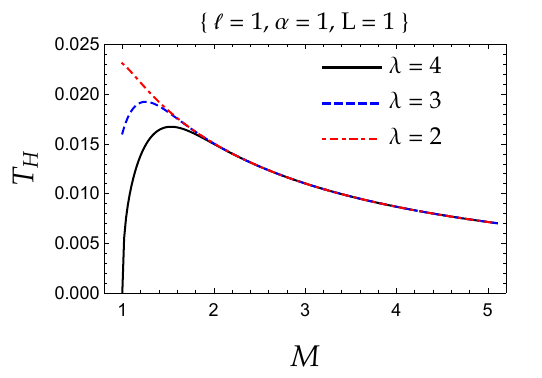}  \
\\
\includegraphics[width=0.32\textwidth]{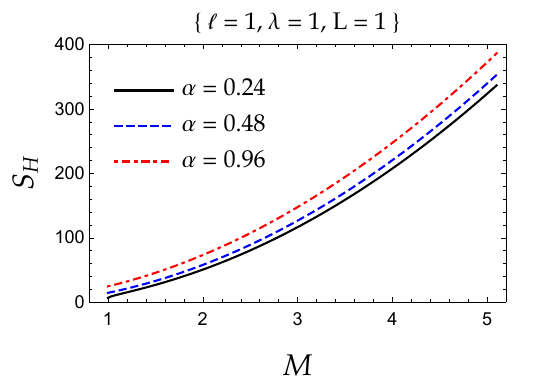}  \
\includegraphics[width=0.32\textwidth]{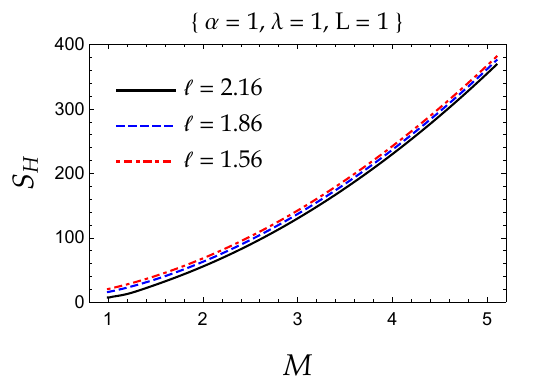}  \
\includegraphics[width=0.32\textwidth]{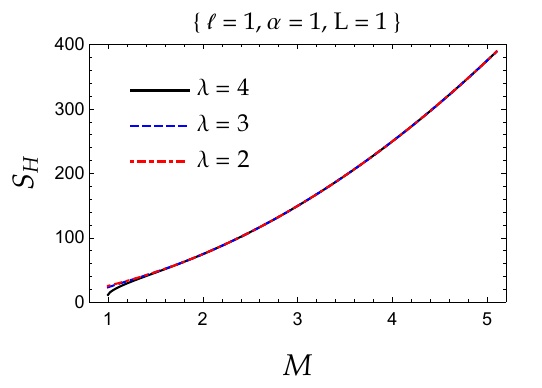}  \
\\
\includegraphics[width=0.32\textwidth]{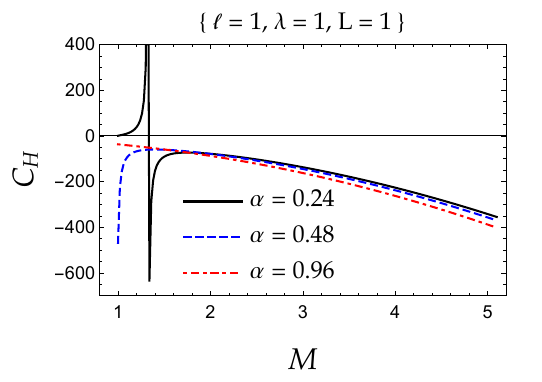}  \
\includegraphics[width=0.32\textwidth]{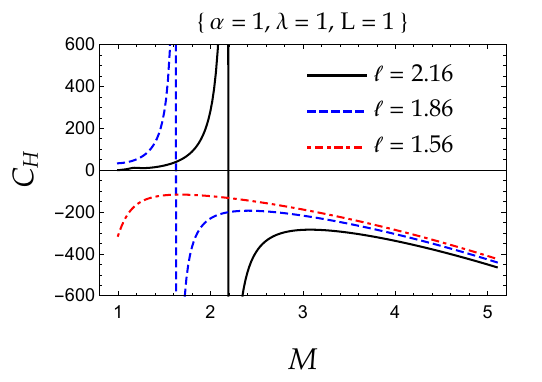}  \
\includegraphics[width=0.32\textwidth]{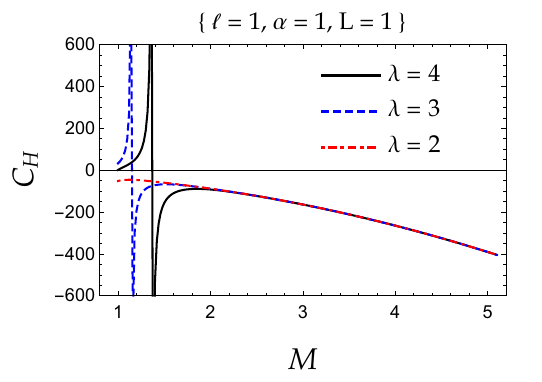}  \
\caption{
{\bf First row:} Event horizon versus mass for the Hairy Hayward regular black hole, for some given parameters $\{l, \alpha,\lambda, L\}$.
{\bf Second row:} Hawking Temperature versus mass for the Hairy Hayward regular black hole, for some given parameters $\{l, \alpha,\lambda, L\}$.
{\bf Third row:} Bekenstein-Hawking entropy versus mass for the Hairy Hayward regular black hole, for some given parameters $\{l, \alpha,\lambda, L\}$.
{\bf Fourth row:} Specific heat versus mass for the Hairy Hayward regular black hole, for some given parameters $\{l, \alpha,\lambda, L\}$.
}
\label{fig:panel-3}
\end{figure*}

Another thermodynamic property under examination is the renowned Bekenstein-Hawking entropy \cite{Gibbons:1976ue}. Such a concept has been generalized and implemented in alternative theories of gravity, in particular, a scalar-tensor theory of gravity. Thus, the general formula is provided by Kang \cite{Kang:1996rj}.
\begin{align}
S &= \frac{1}{4} \oint  {\mathrm {d}^2}x \frac{\sqrt{h}}{G(x)}.
\end{align}
As usual, $h_{ij}$ is the induced metric at the horizon, and $G(x)$ is Newton's coupling, which, in this case, is a constant value. Taking advantage of the symmetry as well as the fact that $G(x) = G(r_H )=G_0=1$ is constant along the
horizon, the above integral takes the form 
\begin{align}\label{eqSs}
S_H &=\frac{\mathcal{A}_H}{4G_0} = \pi r_H^2
\end{align}
Notice that see Fig.~\eqref{fig:panel-3} for details.

%
%

\subsubsection{Specific Heat}
Much emphasis is placed on the different specific heats of thermodynamic systems. Although they can be defined per unit mass, it is considered more appropriate in this context to consider the total heat capacity, which takes the simple form:
\begin{align}\label{eqSH}
C_x &=T \Bigg( \frac{\partial S}{\partial T}  \Bigg) \Bigg|_{x}
\end{align}
where $x$ represents some set of parameters that are assumed to be constant.
Thus, as can be inferred at this point, by setting some parameter as constant we can obtain different specific heat. In such a sense, on the horizon, the heat capacity can be defined as
\begin{align}\label{eqSH2}
C_H &\equiv T_H \Bigg( \frac{\partial S}{\partial T}  \Bigg) \Bigg|_{r_H} 
=
T_H \Bigg( \frac{\partial S}{\partial r_H}  \Bigg) \Bigg( \frac{\partial T}{\partial r_H}  \Bigg)^{-1}.
\end{align}
Albeit the full expression is quite long, we will include it to maintain the discussion self-contained and to include the main thermodynamic properties. Thus, defining $C_H \equiv - \mathbb{N}_1/\mathbb{D}_1$, we have:
\begin{align}
\begin{split}
    \mathbb{N}_1 
    = 
    \ \
    & 
    2 \pi r_H^2 \left(2 l^2 M + r_H^3\right)
    \Bigg[
    4 l^4 M^2 r_H + 4 l^2 M r_H^3 (r_H-3 M) + r_H^7
    \ \ +
    \\
    &
    3 \left(2 l^2 M + r_H^3\right)
    \Bigl(2 l^2 M (r_H-\alpha  L) + r_H^3 (-\alpha  L-2 M + r_H)\Bigl) \times
    \\
    &
    \ln \left(\frac{\alpha  L \left(2 l^2 M + r_H^3\right)}{r_H^3 (\alpha  L+2 M - r_H)-2 l^2 M (r_H-\alpha  L)}\right)
    \Bigg]
\end{split}
\end{align}
\begin{align}
\begin{split}
\mathbb{D}_1 
    = 
    \ \
    & 
    2 r_H 
    \Bigl( 12 l^4 M^2 r_H^3 + 8 l^6 M^3+6 l^2 M r_H^5 (r_H-6 M) + r_H^9 
    \Bigl) +
    \\
    \ \
    &
    9 \left(2 l^2 M + r_H^3\right)^2 
    \Bigl[
    2 l^2 M (r_H - \alpha  L) + r_H^3 (- \alpha  L-2 M + r_H)
    \Bigl] \times
    \\
    \ \
    &
    \ln^2 \Bigg[\frac{\alpha  L \left(2 l^2 M + r_H^3\right)}{r_H^3 (\alpha  L+2 M - r_H)-2 l^2 M (r_H - \alpha  L)}\Bigg]
\end{split}
\end{align}
%

%
\subsection{Hairy Bardeen regular black hole}
%
\subsubsection{Event horizon}
%
Following the same procedure for the previous case, we will solve $f(r_H)$ numerically given the complexity of the lapse function. Thus, for simplicity, we will summarize our main finding in several figures varying the parameters of the model. In concrete, the lapse function to be used in this subsection is (see Eq.~\eqref{eq:bardeendym} for further details):
\begin{align}
    f(r) & = 1-\frac{2 M \,r^{2}}{(E^2 +r^{2})^{3/2}}  - \frac{\alpha L}{r}\bigg(1-e^{-\frac{r^{3}}{\alpha L \lambda^{2}}}\bigg) ,
\end{align}
and the corresponding mass function $m(r)$ can also be defined via the conventional relation
\begin{align}
f(r) &= 1 - \frac{2 m(r)}{r},
\end{align}
taking $m(r)$ the concrete form
\begin{align}
m(r) &= \frac{M r^3}{(E^2 + r^2)^{3/2}}+\frac{1}{2} \alpha  L \bigg(1-e^{-\frac{r^3}{\alpha  \lambda ^2 L}}\bigg).
\end{align}
Note that when $\alpha$ is taken to be zero, we recover the well-known Bardeen black hole solution. Also, after comparing the Hairy-Hayward regular black hole and the present one, we observe that it is possible to go from one solution to the other after and adequate identification of the parameters. 

In this context, we will rename the parameter $E$ as $Q$ in the following, to keep in mind that this solution can be found in the light of nonlinear electrodynamics.

Thus, if we take $Q \rightarrow Q(r,l,M)$ we are able to obtain the Hairy Hayward regular black hole.
As can be observed, the corresponding lapse function depends, in principle, on five independent parameters: $\mathcal{O}:=\{ M, Q, \alpha, L, \lambda \}$, however, as we pointed out in the previous case, the combination $\alpha L$ always appears together, the reason why we can study the effect of the product or just vary one of them keeping the other value equal to one. Thus, the lapse function really depends on the set $\mathcal{O}:=\{ M, Q, \alpha L, \lambda \}$. For a better visualization, we will show how the lapse function evolves by cyclically varying the parameters.
%
\subsubsection{Temperature}
%
As in the previous case, we will analyze the thermodynamics of the second
configuration, so we will introduce the Hawking temperature defined as follows
\begin{align}
    T_H &= \frac{1}{4\pi} 
    \Bigg|
    \frac{\mathrm{d}f(r)}{\mathrm{d}r}\Bigg|_{r=r_H}
    \Bigg|.
\end{align}
The lapse function for this second case is similar to the previous one, and, unfortunately, has the same problem, i.e., the Hawking temperature acquires a relatively complicated form. Removing the $\lambda$-dependence, the concrete expression is then
\begin{align}
T_H &= \frac{1}{4 \pi  r_H} 
\Bigg[
1 - \frac{6 M Q^2 r_H^2}{\left(Q^2+r_H^2\right)^{5/2}}
+
\left( 3 - \frac{3 \alpha  L}{r_H} - \frac{6 M r_H^2}{\left(Q^2+r_H^2\right)^{3/2}} \right) 
\ln \left(\frac{\alpha  L \left(Q^2+r_H^2\right)^{3/2}}{ 2 M r_H^3 - \left(Q^2+r_H^2\right)^{3/2} (r_H - \alpha  L) } \right)
\Bigg].
\end{align}
Be aware and notice that we have expressed the temperature in such a way that when $Q$ and $\alpha$ tend to zero, the standard temperature $T_0 \equiv 1/(4\pi r_H)$ is recovered.

\subsubsection{Bekenstein-Hawking entropy}
\begin{figure*}[ht]
\centering
\includegraphics[width=0.32\textwidth]{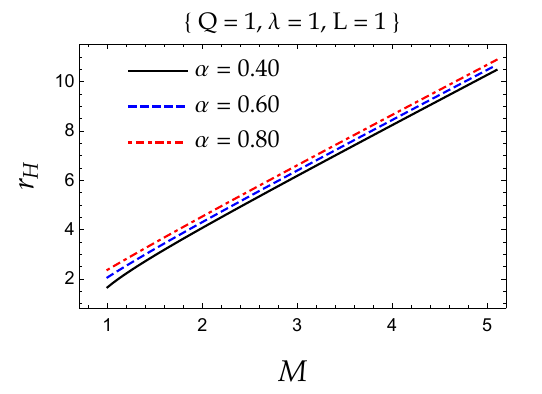}  \
\includegraphics[width=0.32\textwidth]{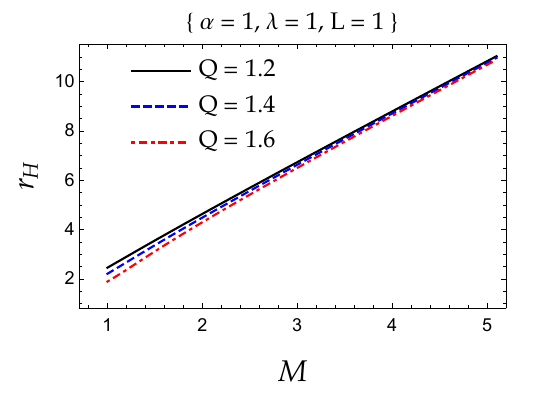}  \
\includegraphics[width=0.32\textwidth]{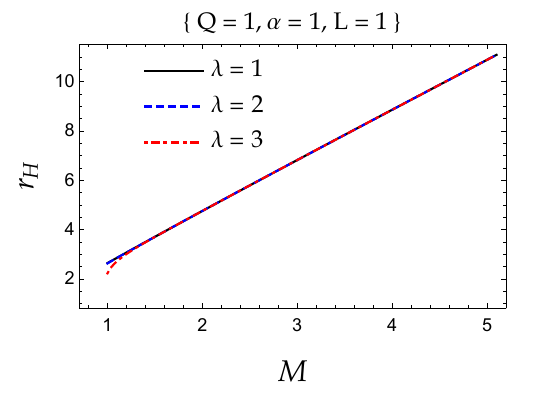}  \
\\
\includegraphics[width=0.32\textwidth]{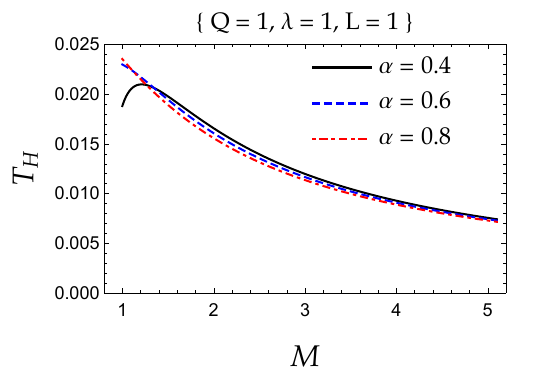}  \
\includegraphics[width=0.32\textwidth]{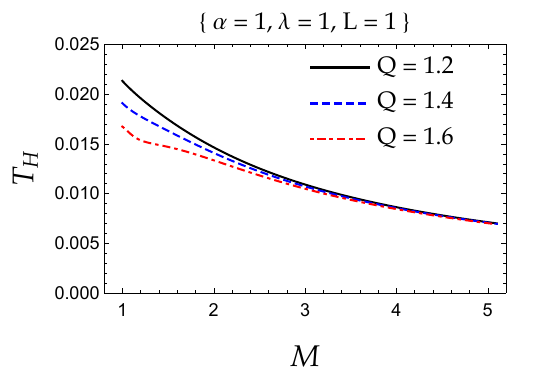}  \
\includegraphics[width=0.32\textwidth]{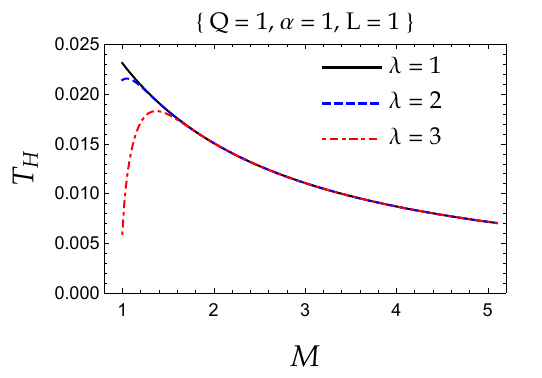}  \
\\
\includegraphics[width=0.32\textwidth]{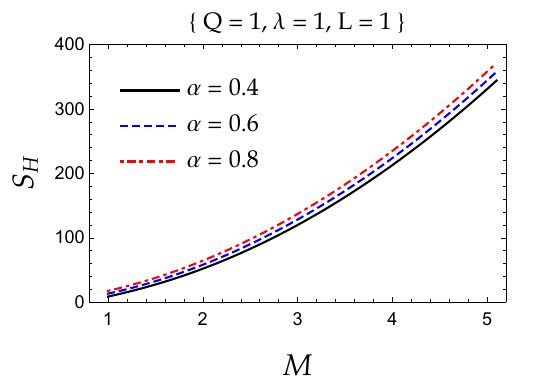}  \
\includegraphics[width=0.32\textwidth]{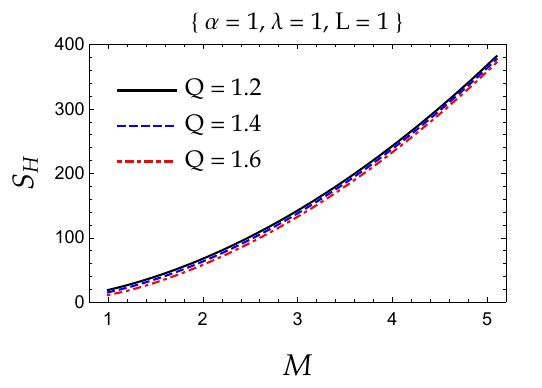}  \
\includegraphics[width=0.32\textwidth]{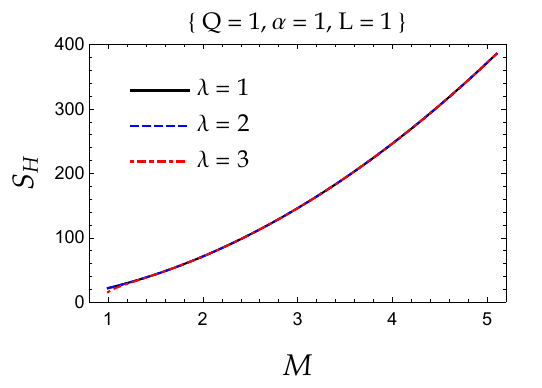}  \
\\
\includegraphics[width=0.32\textwidth]{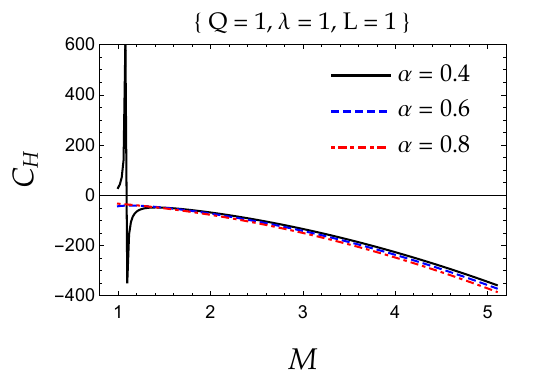}  \
\includegraphics[width=0.32\textwidth]{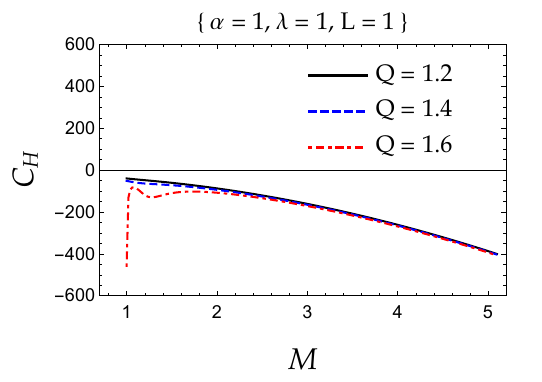}  \
\includegraphics[width=0.32\textwidth]{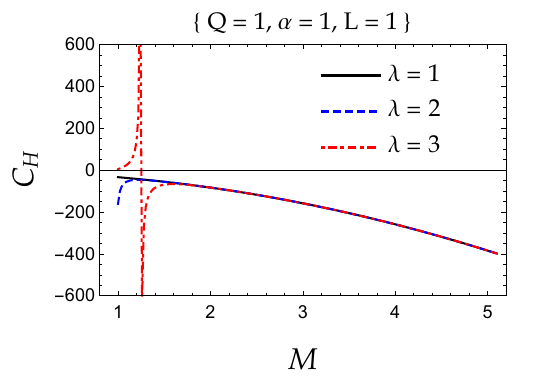}  \
\caption{
{\bf First row:} Event horizon versus mass for the Hairy Bardeen regular black hole, for some given parameters $\{Q, \alpha,\lambda, L\}$.
{\bf Second row:} Hawking Temperature versus mass for the Hairy Bardeen regular black hole, for some given parameters $\{Q, \alpha,\lambda, L\}$.
{\bf Third row:} Bekenstein-Hawking entropy versus mass for the Hairy Bardeen regular black hole, for some given parameters $\{Q, \alpha,\lambda, L\}$.
{\bf Fourth row:} Specific heat versus mass for the Hairy Bardeen regular black hole, for some given parameters $\{Q, \alpha,\lambda, L\}$.
}
\label{fig:panel-4}
\end{figure*}

%
In this subsection, we will just reinforce that the expression required to obtain the Bekenstein-Hawking entropy, in spherical symmetry and in four-dimensional spacetime maintains the same form as in the previous case, namely:
\begin{align}
    S_H &= \frac{\mathcal{A}_H}{4G_0} = \pi r_H^2 .
\end{align}
As always, to determine the event horizon of the hairy Bardeen regular black hole, we solve for the radius at which the radial component of the metric tensor vanishes. Once we have obtained the expression for the event horizon radius, we can then compute the area of the event horizon, and plug it into the Bekenstein-Hawking entropy. As mentioned earlier, the horizon can only be found numerically, which is why we show our results in figures (see Fig. \eqref{fig:panel-4}).

\subsubsection{Specific Heat}
The specific heat is identically defined as in the first case, and removing the $\lambda$-dependence, we can compute to obtain a compact expression which is:
\begin{align}
    C_H &= 
    - \pi r_H^2 
\left[
    \frac{ 
    1 - \frac{6 M Q^2 r_H^2}{\left(Q^2+r_H^2\right)^{5/2}} 
    -
    \left(
    \frac{3 \alpha  L}{r_H} + \frac{6 M r_H^2}{\left(Q^2+r_H^2\right)^{3/2}} - 3 
    \right) 
    \ln 
    \left(
    \frac{\alpha  L \left(Q^2+r_H^2\right)^{3/2}}{2 M r_H^3 -\left(Q^2+r_H^2\right)^{3/2} (r_H-\alpha  L) }
    \right) 
    }{
    1 - \frac{15 M Q^2 r_H^4}{\left(Q^2+r_H^2\right)^{7/2}}  +
    \left(
    \frac{9 \alpha  L}{2r_H} + \frac{9 M r_H^2}{\left(Q^2+r_H^2\right)^{3/2}} - \frac{9}{2} \right) 
    \ln ^2
    \left( 
    \frac{\alpha  L \left(Q^2+r_H^2\right)^{3/2}}{2 M r_H^3 -\left(Q^2+r_H^2\right)^{3/2} (r_H-\alpha  L)}
    \right)  }
\right]
\end{align}
Notice that $C_H$ has a global negative sign which, in principle, suggests the black hole is unstable. However, as in the previous case, in practice, such stability/instability depends on a non-trivial way of the combination of the parameters of the model and also the deviations from the simplest case (i.e., when $Q$ and $\alpha$ are taken to be zero). 

%

\section{Conclusion}
Nowadays, gravitational decoupling is a powerful tool to explore new solutions of Einstein's field equations with more realistic matter content. New matter content causes geometrical deformation that leads to new descriptions of known objects. In this research, we have considered two well-known solutions of the Einstein field equations describing regular Hayward and Bardeen black holes as seed spacetimes. By considering different matter contents, we have obtained several new non-singular black hole solutions of the Einstein equations. However, we focus only on two of them, representing deformed Hayward \eqref{eq:dymhay} and deformed Bardeen \eqref{eq:bardeendym}, non-singular black holes with de Sitter core and Schwarzschild-like behavior in infinity. 
We then studied some thermodynamical properties of the newly obtained solutions and localized their event horizons. 
In particular, we have calculated 
i) the event horizon, which grows as the black hole mass increases and is required to evaluate the remaining thermodynamic properties,
ii) the temperature, which decreases for large values of the mass $M$ and exhibits a maximum depending on the parameters involved and the mass range,  
iii) the Hawking entropy, which increases with increasing $M$
and finally
iv) the specific heat, $C_H$, which is a thermodynamic quantity that indicates whether the black hole is stable or not. If $C_H < 0$, the black hole is unstable. Our results show that our solutions are unstable for moderate and large values of the black hole mass.
We also found that some minimal geometric deformations can lead to the formation of singularities, or the solution does not have a de Sitter core in the center (which can indicate a weak energy violation) and does not behave like a Schwarzschild solution in infinity. 
The nature of this geometric deformation could be the presence of an accretion disk around a black hole. By considering the system black hole versus accretion disk as a single system described by the special spacetime, we can study the influence of the accretion process on different properties of a black hole.

\section*{Acknowledgments}
A. R. acknowledges financial support from the Generalitat Valenciana through PROMETEO PROJECT CIPROM/2022/13.
A. R. is funded by the María Zambrano contract ZAMBRANO 21-25 (Spain) (with funding from NextGenerationEU).
The work was performed as part of the SAO RAS government contract approved by the Ministry of Science and Higher Education of the Russian Federation.
M. M. gratefully acknowledges partial support from the Theoretical Physics and Mathematics Advancement Foundation BASIS, grand no.20-1-5-109-1.

\end{document}